\documentclass{kluwer}
\usepackage{epsfig}
\usepackage{graphicx}
\newcommand{\arcsec}{$^{\prime\prime}$}
\begin{document}

\begin{article}
\begin{opening}

\title{ RHESSI Images and spectra of two small flares}

\author{L. \surname{Maltagliati}}
\author{ A. \surname{Falchi}}
\institute{Osservatorio Astrofisico di Arcetri, L. Fermi 5, I-50125
Firenze, Italy}
\author{L. \surname{Teriaca}}
\institute{Max-Planck-Institut f\"{u}r Sonnensystemforschung 
Max-Planck-Str. 2, 37191 Katlenburg-Lindau, Germany}
\begin{abstract}
We studied the evolution of two small flares (GOES class C2 and C1)
that developed in the same
active region with different morphological characteristics: one is
extended and the other is compact. We analyzed the accuracy and the
consistency of different algorithms implemented in RHESSI software
to reconstruct the image of the emitting sources, for energies between 3
and 12~keV. We found that all tested algorithms give consistent results
for the peak position while the other parameters can differ at most by a
factor 2. {\it Pixon} and {\it Forward-fit} generally converge to 
similar results but {\it Pixon} is more reliable for reconstructing a
complex source. We investigated the spectral characteristics of the two flares
during their evolution in the 3~--~25~keV energy band. We found that 
a single thermal model of the photon spectrum is inadequate to fit the 
observations and we
needed to add either a non-thermal model or a hot thermal one.
The non-thermal and the double thermal fits are comparable.
If we assume a non-thermal model, the non-thermal energy is always higher
than the thermal one.
Only during the very final decay phase a single thermal model fits
fairly well the observed spectrum.
\end{abstract}
\end{opening}

\section{Introduction}
 Since February 2002, the RHESSI (Reuven Ramaty High-Energy Spectroscopic Imager)
satellite carries out
X-ray and $\gamma$-ray observations of flares and other forms of solar
magnetic activity \cite{Linetal:02}.
In particular, the 3~--~25~keV spectral band is of great relevance.
It is comprehensive of the thermal and non-thermal domain of the
bremsstrahlung, giving us information about the energy deposition of the
particles accelerated in the impulsive phase. The low-energy non-thermal
electrons are especially crucial in small events and in microflares.
Besides, a broadened emission line feature at around 6.7~keV, mainly due 
to Fe ions, and a weaker line feature 
at around 8~keV due to Fe and Ni ions are observable and can give information on
electron distributions different from the ones obtained from the
bremsstrahlung continuum.
The X-ray band 3~--~25~keV is observed by RHESSI with unprecedented
spectral resolution (1~keV) and imaging capability (2.3\arcsec). 
Moreover, when both attenuators are out, RHESSI has a much larger
effective
area than any earlier instrument, providing information on low-level
energy releases with much higher sensitivity.

These RHESSI characteristics allow extensive studies of microflares
related to both imaging and spectra (among others: \opencite{KruckerEtal02}, 
\opencite{Benz&Grigis02}, \opencite{Kruckeretal:04},
\opencite{Hannahetal:04}, \opencite{Liuetal:04}).
Images at low energy of these small events are generally reconstructed
using the {\it Clean} algorithm available in RHESSI software
(\opencite{HurfordEtal02}, \opencite{SchwartzEtal02}). 
In fact, in this process of image reconstructions, {\it Clean}, together
with {\it Pixon} \cite{MetcalfEtal96} and {\it Forward-fit}
\cite{AschwandenEtal02}, converge to the best solutions
for a simple source morphology 
 and the photometric accuracy seems to be
uniform in different energy ranges higher than 10~keV
\cite{AschwandenEtal04}. 
It may be of interest
to test the photometric accuracy and the consistency of the image
reconstruction algorithms for more complex sources and lower energies.

It seems quite established that at the peak time of a flare 
the observed spectra up to $\approx 10$~keV can be fitted with a single thermal 
model (T $\approx 10$~MK) of the photon spectrum. At higher energies a good fit can be 
obtained adding a non-thermal bremsstrahlung
emission with energy cutoff below 10~keV and  a power-law index
$\gamma$ between 5 and 8, meaning that
the spectra of small flares are usually softer than the ones of large
flares. \inlinecite{Benz&Grigis02} obtained a good fit of a microflare
spectrum at the maximum phase also adding a second thermal component
with higher temperature (T$\approx 25$~MK), but they reject this
interpretation because this hotter component is no more present during
the cooling and then the behaviour of the photons higher than 10~keV is
more consistent with the standard flare scenario of precipitating
non-thermal electrons.
During the decay phase instead, the spectra can be fitted
either with a thermal component only \cite{Liuetal:04} or with a
non-thermal component with an index $\gamma\approx10$
(\opencite{KruckerEtal02}, \opencite{Hannahetal:04}). Therefore it is
very important to study the behaviour of the spectra of small flares
during their evolution in order to investigate if the hardness of the spectrum
changes with time, as can be observed in larger flares that show a
direct correlation between the hard X-ray flux and the spectral hardness
(\opencite{Fletcher:02}, \opencite{Hudson:02}, \opencite{Grigis:04}).

In this paper we present a study on the morphological and spectral properties
of two small flares (GOES class C2 and C1) in the energy range
3~--~25~keV. In particular we compare the
image reconstruction algorithms to carefully check the characteristics of
these two events
at these low energies, when both attenuators are out, and we investigate
how their X-ray spectrum evolves with time.

\section{Observations}
An observing campaign coordinated between ground and space based 
instruments was planned to observe flare events, sampling the solar
atmosphere from the chromosphere to the corona. Monochromatic 
images at several wavelengths, in the continuum and within the 
H$\alpha$ line have been acquired by means of the tunable Universal 
Birefringent Filter (UBF) and the Zeiss filter at the
Dunn Solar Tower of the National Solar Observatory (NSO) / Sacramento Peak. 
The field of view (FOV) is of about 150\arcsec$\times$150\arcsec\ with a spatial scale
of 0.5\arcsec$\times$0.5\arcsec\ and a temporal cadence of few
seconds.
Spectra have been acquired with the Horizontal Spectrograph in 3
chromospheric lines (Ca~{\sc ii}~K, He~{\sc i}~D3 and H$\gamma$), with a temporal
cadence of 4~s. 
Simultaneously, spectroheliograms of the same region were
obtained in transition region and coronal lines
with the Coronal Diagnostic Spectrometer (CDS) aboard SOHO
\cite{HarrisonEtal95}. 
During this campaign we
observed the region NOAA AR 10061 (N10W20) on 2002 August 11 from 14:00 to 19:00 UT.
Two homologous flares (GOES class C2 and C1) developed in this region
around 14:40~UT and 16:25~UT. 
 In this paper we mainly concentrate on the
diagnostic provided by RHESSI in conjunction with H$\alpha$ data. The
flare dynamics, using also spectra from NSO and CDS, will be analyzed in a
later paper. 

\begin{figure}
\centerline{\includegraphics[width=7cm]{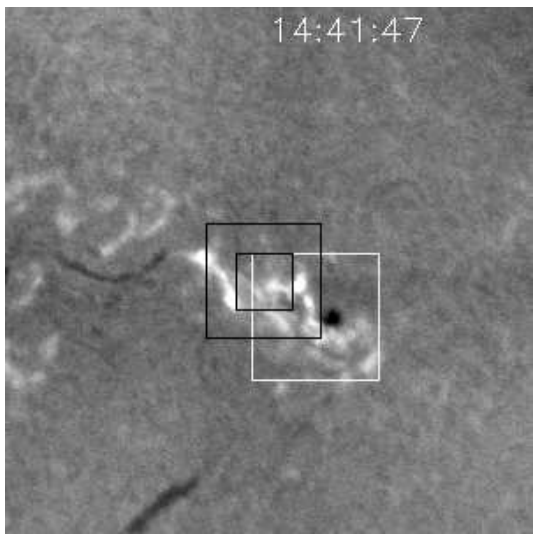}}
\caption{Image of the active region NOAA AR 10061 acquired by the HASTA
telescope in the blue wing of H$\alpha$,  approximately at the peak
time of the first flare (West at right, North at top). Two bright
ribbons are clearly visible. The white box 
(150\arcsec $\times$ 150\arcsec), that includes only one ribbon,
indicates the FOV of the UBF/National Solar Observatory. The black boxes
indicate the FOV used for RHESSI image reconstruction (see Sec. \ref{ima:rec}).}
\label{ima:hasta}
\end{figure}

The image of the whole active region at 14:41:47~UT (Figure \ref{ima:hasta}), 
acquired with the
 HASTA telescope (San Juan, Argentina) in the blue wing of H$\alpha$ line, 
shows the two ribbons  developed during  the first flare. Unfortunately the FOV 
of the UBF (indicated in Figure~\ref{ima:hasta} by the white box) includes 
only the western ribbon of the flare. 

\begin{figure}
\centerline{\includegraphics[width=12cm]{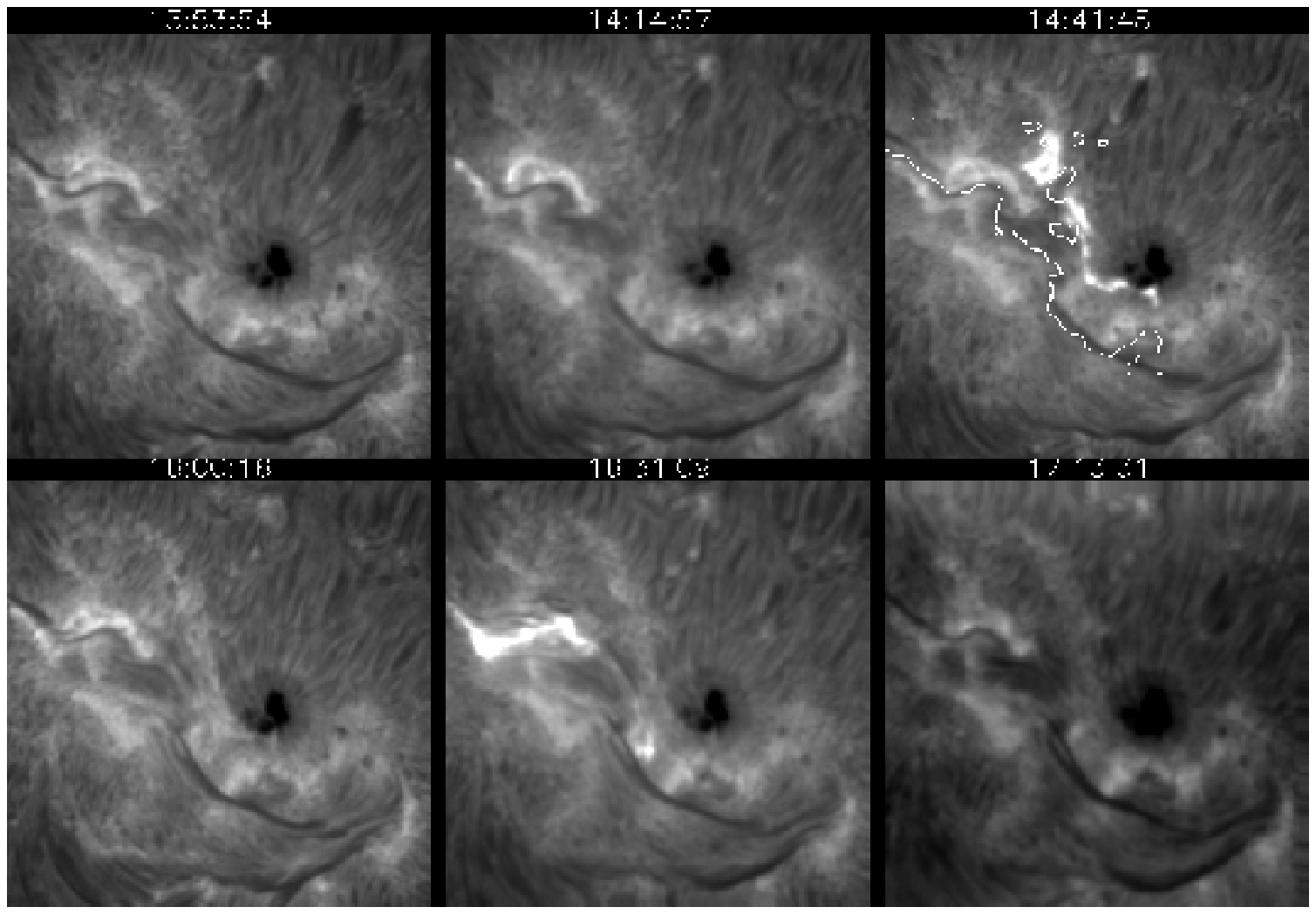}}
\caption{Evolution of the two flares. Images have been acquired with
UBF at the center of H$\alpha$ line. In the upper right panel the white
line indicates the apparent magnetic neutral line obtained from MDI/SOHO data. 
At 14:41:48, the time of the
maximum of the first flare, only one ribbon is within the FOV (see 
Figure~\ref{ima:hasta}). At 16:31:09~UT, the peak time of the second flare, the
ribbons follow the underlying filament and appear very close to each
other.}
\label{ima:ubf}
\end{figure}

In Figure~\ref{ima:ubf} we show 
the high resolution images obtained with
UBF in the H$\alpha$ line center during the two flares.
Both flares 
have the characteristics of eruptive two-ribbon flares and are indeed 
triggered by the eruption of different portions of the same filament. 
However  they exhibit very different morphologies. 
The upper panels show that the upper eastern part of the filament 
disappears  at 14:14:57~UT.  The emission maximum 
of the western ribbon is reached at 14:41:48~UT when two extended parallel ribbons 
developed in NE-SW direction (see Figure~\ref{ima:hasta}),  
at the two sides of the filament. MDI/SOHO \cite{ScherrerEtal:95} 
full disk magnetogram taken at 14:27:34~UT is used to determine the
position of the apparent magnetic neutral line and the alignment, done using
MDI and UBF continuum, is estimated to be within 1\arcsec. The 
neutral line follows very well the filament and the ribbon positions hint 
at a system of large coronal loops ($\approx 70\,000$~km) in East-West direction. 
Bottom panels show that at 16:00:16~UT
the emission from the first episode has almost disappeared and the whole
filament is visible. At 16:31~UT 
 very compact ribbons develop nearby the central part of the filament,
 suggesting the presence of stressed and lower loops. At 17:13:31~UT the
 filament structure is again in the pre-flare conditions.

\begin{figure}
\centerline{\includegraphics[width=12cm]{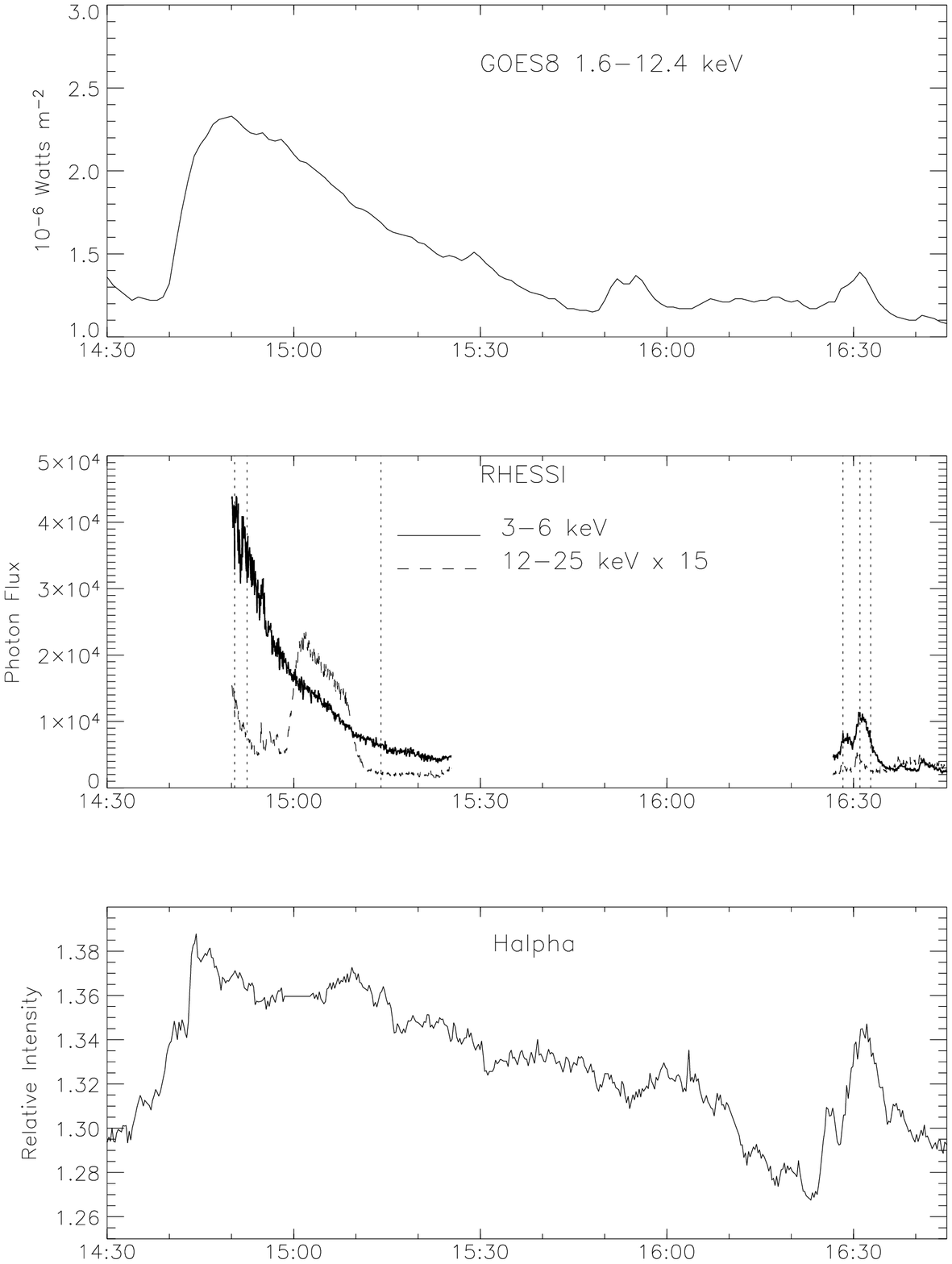}}
\caption{Light curves of the two flares observed by GOES (top), RHESSI
(middle) and UBF (bottom).
In the middle panel, the vertical dotted lines indicate the times at which 
both the RHESSI images and spectra have been studied. 
Notice the strong increase between 14:56~UT - 15:12~UT in the
energy bin 12~--~25~keV that is due to a particle precipitation event (see
text).} 
\label{crv:all}
\end{figure}

Figure \ref{crv:all} shows the light curves of the two events observed in soft
X-rays at 1~--~8 \AA \ by GOES, in the energy bins 3~--~6 and 12~--~25~keV by
RHESSI and in H$\alpha$ line center by UBF.
We notice that GOES recorded a C1 flare at about 15:50~UT which did not occur
in the observed active region, and that GOES and UBF light curves are
very similar for the two studied flaring episodes.
RHESSI 
was in its night during the impulsive phase of the first event and
acquired data only during its decay phase. 
For the second event RHESSI
exited from its night just after the very first spike at 16:25:54~UT, 
well visible in the H$\alpha$  curve, missing the beginning
of the impulsive phase. 
The significant increase of the counts in the time
interval 14:56~UT - 15:12~UT, at energies higher than 10~keV, is due to
charged particle precipitation event and then the data in this time
interval cannot be considered.
RHESSI had both thin and thick attenuators out for the whole time
interval.

\section{RHESSI Image Reconstruction}\label{ima:rec}

In RHESSI software 5 different algorithms have been implemented to
reconstruct  images from the `back projection' map, which
contains sidelobes due to the modulation pattern of the collimators.
\inlinecite{AschwandenEtal04} carried out a comparative study on the 
accuracy of the algorithms between 10 and 60~keV for a strong flare
with two resolved sources.
We did analyze the efficiency and the accuracy of different algorithms in
the energy range 3~--~25~keV when both thin and thick attenuators are
out and for two different source morphologies, i.e.: an extended and a
compact flare. 
This energy range is crucial for the understanding of
small flares  and it might present different problems because of 
instrumental performances.

\subsection{Extended flare}\label{ima:rec:ext}
The minimum FOV to include all features of this 
complex-source flare is 128\arcsec$\times $128\arcsec\ (larger
black box in Figure~\ref{ima:hasta}),
centered on the point with solar coordinates (340,80), 
which seems to be the center of the X-ray emission.
We reconstructed the images  with a pixel size of 2\arcsec$\times$ 2\arcsec\
considering a time interval  
$\Delta t = 60$~s which is short enough to give a low flux gradient 
at the considered time.
We set off the time variability option (see later). 
The choice of collimators that contributed usefully to the definition
of images has been accurate even if there is a certain arbitrariness.
 
Collimator 2 cannot be employed for imaging below 20~keV due to a
failure during the early phases of the mission \cite{Smithetal:02}.
We made separate back-projection images for each other collimator and
discarded those which show no obvious source. At the end,
we decided to use collimators 5~--~8 (FWHM of the lower grid =
20.4\arcsec).
Collimator 9 (FWHM of 186\arcsec) is excluded,
as its low resolution does not add any new features, given the field of
128\arcsec. 
 Collimator 7 is also excluded in the 3~--~6 keV range
because its energy threshold is 7~keV and its resolution is 3~keV.
The reduced number of detectors used (only 3 over 9 in the 3~--~6 keV
range) could raise concerns
on the results of this comparative analysis. However, we will 
show in the next Sections that the source parameters are not strongly
affected by the number of detectors used in the reconstruction process.

\begin{figure}[t]
\centerline{\includegraphics[width=12cm]{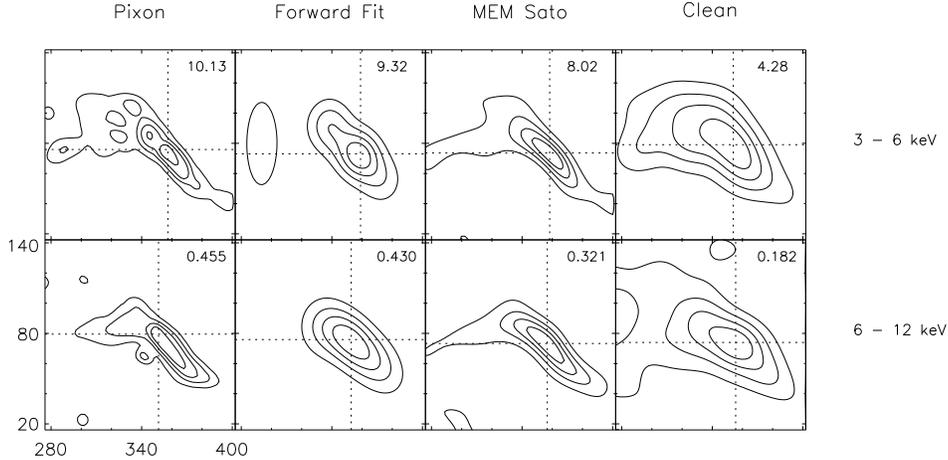}}
\caption{Extended flare: isocontours (20~\%, 40~\%, 60~\%, 80~\% of the peak 
intensity, indicated
in the upper right corner) of the images reconstructed 
with four algorithms at 14:50:30~UT.}
\label{imag:exten}
\end{figure}
We compare the results obtained using four imaging algorithms available
in the RHESSI data analysis software ({\it Pixon, Forward-fit, 
MEM-Sato, Clean}).
We tried also the {\it MEM-Vis} algorithm but we could not obtain 
the convergence for this method.
Due to the complexity of the source, to obtain a good
reconstructed image with {\it Forward-fit} we had to use 4 elliptical Gaussians. 

We considered the three energy ranges 3~--~6, 6~--~12 and 12~--~25 keV, but 
we had to exclude from our analysis the latter range, because 
the pileup effect is important for this flare (see Sect.~\ref{spe:pile}).
No correction for pileup is currently available for 
imaging, because the magnitude of the pileup varies on the same short 
timescales as the modulation, thus an exact correction for pileup would 
require an a priori 
knowledge of the image at all energies (Hurford, private communication).

In Figure~\ref{imag:exten} we
show the contours of the images obtained with the different 
algorithms in the 3~--~6 and 6~--~12 keV range at 14:50:30~UT (all
RHESSI times refer to the center of the adopted integration interval).
The other selected times provide very similar results and are not
presented here.

\begin{table}%[t]
\caption{Parameters obtained with different algorithms for the main
source of the extended flare at 14:50:30~UT (see text for explanations).} 
\label{tab:exten}
\begin{tabular}{lccccccccc} \hline
Algorithm & Energy & x$_p$,y$_p$ & F$_p$ & F$_{tot}$
& w$_x$& w$_y$ & H/C \\ 
\hline
Pixon& 3-6 & 357.4,75.9 & 10.1 & 17362 &  35.0&20.9&0.30 \\
Forw. Fit& 3-6 & 359.0,73.0 & 9.3 & 18240 & 38.2&32.2&0.19 \\
MEM Sato& 3-6 & 358.3,73.6 &  8.0 & 16880 & 36.0&24.4&0.38 \\
Clean& 3-6 & 353.9,79.1 & 4.3 & 15000 & 64.1&46.8& 0.24 \\ \hline
Pixon& 6-12 & 351.2,79.5 & 0.45 & 757.87 &  36.0&21.4&0.16 \\
Forw. Fit& 6-12 & 352.8,75.9 & 0.43 & 673.26 & 37.4&32.9& 0.05 \\
MEM Sato& 6-12 & 356.7,73.3 & 0.32 & 735.80 & 35.9&29.2& 0.26 \\
Clean& 6-12 & 355.5,74.0 & 0.18 & 678.92 & 60.0&42.9& 0.25 \\ \hline
\end{tabular}
\end{table}
For each reconstructed image we give in Table~\ref{tab:exten} the
parameters obtained in the 3~--~6 and 6~--~12~keV energy ranges, 
that might help to
assess the stability and the congruity of the different methods:

\begin{itemize}
\item The peak position x$_p$ and y$_p$ (arcsec), found by a parabolic fit 
of the nine pixels
 surrounding the image peak, and the peak flux on the detector F$_p$ 
(photon~s$^{-1}$~cm$^{-2}$~arcsec$^{-2}$).
\item The total flux over the image on the detector F$_{tot}$ 
(photon~s$^{-1}$~cm$^{-2}$).
\item The equivalent widths $w_{x}$ and $w_{y}$ along the x and y direction 
give an indication of the linear dimension in arcsec and are defined
following \inlinecite{AschwandenEtal04}. 
\item The $[Halo]/[Core]$ flux ratio H/C, where $[Core]$ is  the
flux per unit surface in the part of the image
inside of a circle centered on the peak emission with a radius 
$\simeq 1.5$ times the resolution of the finest grid used, and 
$[Halo]$  is the
flux per unit surface in the part outside of it \cite{AschwandenEtal04}.
\end{itemize}

The peak positions agree within few seconds of arc, $\approx$ 1-2~px, and the
total fluxes agree within 20 \%. 
 The peak flux values and the linear dimensions $w_{x,y}$ differ more
than a factor of 2: {\it Pixon} and {\it Forward-fit} give similar
results while
{\it Clean} finds the lowest value for the peak flux and the largest for
the linear dimensions.  This is
however an expected result, because {\it Clean} does not
deconvolve the  source from the Point Spread Function (PSF) of the
instrument. 
However, the flux integrated over the source ($F \approx
F_{p}~w_{x}~w_{y}$) differs up to a factor of 1.5 and is not strictly 
conserved for different algorithms as in the case discussed 
by \inlinecite{AschwandenEtal04}.

The $[Halo]/[Core]$ flux ratio indicates how much each 
method spreads  the photons detected over the FOV outside the real sources.
However, we stress that, especially in the case of a multiple-source
flare, the
$[Halo]$ takes into account also photons coming by other real sources present
in the FOV.
{\it Forward-fit}, as implicit in the hypotheses used to reconstruct the
image, obtains the lowest value for this ratio. 
{\it MEM-Sato} finds a source
size comparable to {\it Forward-fit} and {\it Pixon}, but it obtains the
highest $[Halo]/[Core]$. Thus, we can say that {\it MEM-Sato} leaves more 
photons out of the source.

Finally, we investigate how the image reconstruction process is affected
by the presence of unmodulated flux due to a combination of scattered flare photons
within the instrument and a non-flare background flux \cite{AschwandenEtal04}.
{\it Pixon} and {\it Forward-fit} have options to detect and remove
this component. The corrected images obtained with {\it Forward-fit} are
very similar to the uncorrected ones giving variations of a few percent
on the parameters reported on Table~\ref{tab:exten}. Instead, the parameters
obtained with {\it Pixon} differ up to 40 \% (if we
exclude the total flux that is almost unchanged)  giving very different results
even in the peak position ($\Delta x_p=-8$\arcsec, $\Delta y_p=+9$\arcsec)
while the general morphology of the emitting region appears to be
unchanged.
 
The {\it Clean} algorithm has the possibility to correct the image for
the residual map: this procedure strongly affects the total flux 
that decreases more than 20 \% and then differs more than 40 \% from values
found with other
algorithms. However, the $[Halo]/[Core]$ ratio decreases as well
as the flux integrated over the source.
This means that the correction for the residual map subtracts photons from
the field of view, but mainly from the halo.

\subsection{Compact flare}\label{ima:comp}
\begin{figure}%[t]
\centerline{\includegraphics[width=12cm]{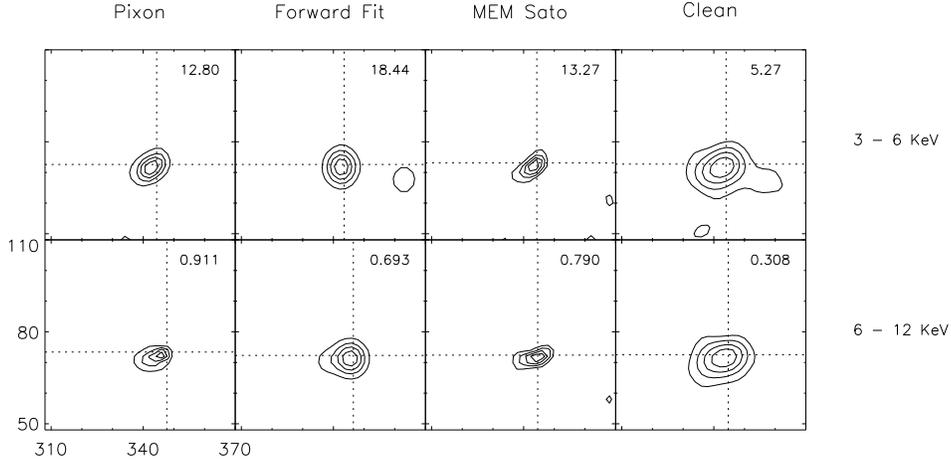}}
\caption{The same as in fig. \ref{imag:exten} at 16:27:58 UT.}
\label{imag:comp}
\end{figure}
To reconstruct the images of this compact event  we
could use a FOV of
64\arcsec$\times$64\arcsec\ (smaller black box in Figure~\ref{ima:hasta}), 
centered on the point with solar coordinates (340,80), and a 
 pixel size of 2\arcsec$\times$2\arcsec.
Since the light curve (Figure~\ref{crv:all}) shows rapid variations with two rise
phases, two maxima and a decay phase, the time interval required
for the image reconstruction should be short and we set  $\Delta t = 20$~s.
To create the images we used collimators 
3~--~7 with a FWHM = 6.8\arcsec of the finest grid considered. 
Collimators 8~--~9 were excluded, because their FWHM (96\arcsec\ and
186\arcsec\ respectively) exceeds the FOV and collimator 7 was excluded
in the range 3~--~6~keV, as in the previous case. 
This flare has a simpler structure and we were able to use 2
elliptical Gaussians to reconstruct images with {\it Forward-fit} algorithm.

We examined the four imaging algorithms  in different evolution phases of
the flare with
the time variability option switched off. 
The general behaviour of the algorithms does not change with the flare
evolution and
in Figure~\ref{imag:comp} we show as an example the images reconstructed 
in the first 
rising phase (16:27:48~UT -- 16:28:08~UT) for the energy ranges 3~--~6 and 
6~--~12~keV. In Table
\ref{tab:comp} the parameters obtained for these images are reported.

Analyzing the parameters given in Table~\ref{tab:comp} we see that the 
peak positions found with all algorithms
agree very well within 2\arcsec,  and the total fluxes 
agree within 10 \%, but only if we exclude the {\it Clean} algorithm. 
 {\it Pixon} and {\it Forward-fit} give very similar results for the peak
flux and the linear dimensions while {\it MEM-Sato} and {\it Clean} give
different values resulting in a flux integrated over the source a factor
of 2 lower.
The image obtained with
{\it Clean} shows actually very high negative residuals out of the main source: 
consequently the total flux over the FOV is much lower than that
obtained with other algorithms and moreover the
$[Halo]/[Core]$ ratio cannot be reliably computed. 
Even if we use the available option  in order to correct the image 
for the residual map we cannot obtain more reliable results.
As in the case of the extended flare the {\it MEM-Sato} algorithm 
gives the highest value of $[Halo]/[Core]$ ratio.
Finally we use for {\it Pixon} and {\it
Forward-fit} the options to remove the unmodulated flux:
in this case the parameters change only of a few percent.

\begin{table}%[t]
\caption{Parameters obtained with different algorithms for the main
source of the compact flare at 16:27:58~UT.} \label{tab:comp}
\begin{tabular}{lccccccc} \hline
Algorithm & Energy & x$_p$,y$_p$ & F$_p$ & F$_{tot}$
& w$_x$& w$_y$ & H/C \\ 
\hline
Pixon& 3-6  & 344.4,72.6 & 12.8 & 2023 & 10.1&9.5& 0.11 \\
Forw. Fit& 3-6 & 343.5,72.8 & 15.0 & 2121 & 10.2&9.9&0.04 \\
MEM Sato& 3-6 & 344.3,73.1 & 13.3 & 1885  & 8.4 & 7.1& 0.21 \\
Clean& 3-6 & 344.0,72.7 & 5.3 & 724 & 17.0&11.5& -- \\ \hline
Pixon& 6-12 & 347.7,73.4 & 0.9 & 77.21 &10.2&9.7& 0.05 \\
Forw. Fit& 6-12 & 346.5,72.3 & 0.7 & 67.37 &10.4&9.7& 0.004 \\
MEM Sato& 6-12 & 344.6,72.4 & 0.8 & 75.13 &9.7&8.8& 0.08 \\
Clean& 6-12 & 344.7,72.5 & 0.3 & 19.15 &15.7&10.2& -- \\
\end{tabular}
\end{table}

For both the extended and the compact flare we did some other tests to
verify the stability of the parameters found. 
Changing the pixel size from 1\arcsec\ to 4\arcsec\ does not change
the parameter values more than 1 \%. A similar variation is
obtained with the time variability option on and off, indicating that if
the time interval considered is short enough to give a low flux
gradient this option can be dropped.
For the compact flare in the energy range 12~--~25 keV, when the counts are
very low, we could not obtain the
convergence of any algorithm if we consider a $\Delta t = 20$~s, but
we should use $\Delta t = 60$~s. Within this interval the counts are
highly variable and to reconstruct an image it is mandatory
to set the time variability option on.

 We can conclude that in our two cases of small flares,
the results obtained by different algorithms do not depend on the
chosen energy band within the 3~--~12 keV range  even if the
detector 7 has been excluded only in the 3~--~6~keV band.
This indicates that the
reconstruction algorithms are reliable even when only few collimators
are useful to detect the emission
source (see Sect.~\ref{ima:rec:ext}).
Different algorithms may find different problems in reconstructing the
images, but the problems seem not to be related to the extension or 
compactness of the source but rather to the complexity of the region. 

 {\it MEM-Sato} always gives the highest value for H/C as found by 
\inlinecite{AschwandenEtal04}.  However, in all
the cases we examined it does not under-resolve the source, giving always
linear sizes of the same order of the other algorithms.
{\it Pixon} and {\it Forward-fit} give similar results but
in the case of the extended flare the images reconstructed by 
{\it Forward-fit} do not seem to show the real complexity of the source. Thus
we use {\it Pixon} to reconstruct the images for the analysis of flare
evolution.

\section{Spectra evolution}

We analysed the spectral characteristics for the spatially integrated
flux of our two flares as they evolve in time.
The count spectrograms were created using the standard RHESSI software 
(updated April 2004). 
We used the front segments of the detectors in the 3~--~25 keV energy range,
with the energy binning set to 1/3~keV, 
excluding detectors 2 and 7, which have lower energy resolution
\cite{SmithBBB, SmithPPP}.
The full spectrum response matrix has been used to calibrate the data, 
and the SPEX code (\opencite{Schwartz:96};
\opencite{Smithetal:02}) has been used for the spectral fitting. We
considered 
energies above 4~keV, because the lower energy intervals are not 
well calibrated yet. We tried to fit the observed count spectrum 
with three different models of photon spectrum: 
\begin{itemize}
\item A single thermal bremsstrahlung emission whose best-fit parameters
are the temperature T of the isothermal emitting plasma and its
emission measure EM.
\item A thermal plus a non-thermal bremsstrahlung emission  
whose best-fit parameters
are, besides T  and EM of the isothermal emitting plasma, the power law
index $\gamma$ and the normalization of the power-law at the
normalization energy fixed at 10~keV. We assume a power-law index fixed
to 1.5 for energy below the energy cutoff, that is determined as the
one that gives the lowest $\chi^2$ value (see Sec. \ref{spe:pha}).
 We recall that $\chi^2$ is the reduced one, computed considering
the numbers of free parameters. 
\item A double thermal bremsstrahlung emission that assumes the presence of 
 two isothermal components and  gives as best-fit parameters the
 temperatures T and T$_h$ and the emission measures EM and EM$_h$, for
 the cold and the hot component, respectively.
\end{itemize}

\subsection{Choosing the background} \label{spe:cho_back}
RHESSI is a high-background instrument and the background selection
in the spectra analysis is crucial in particular for small and low-energy
flares. 
\begin{figure}%[t] 
\centerline{\includegraphics[width=12cm,height=7cm]{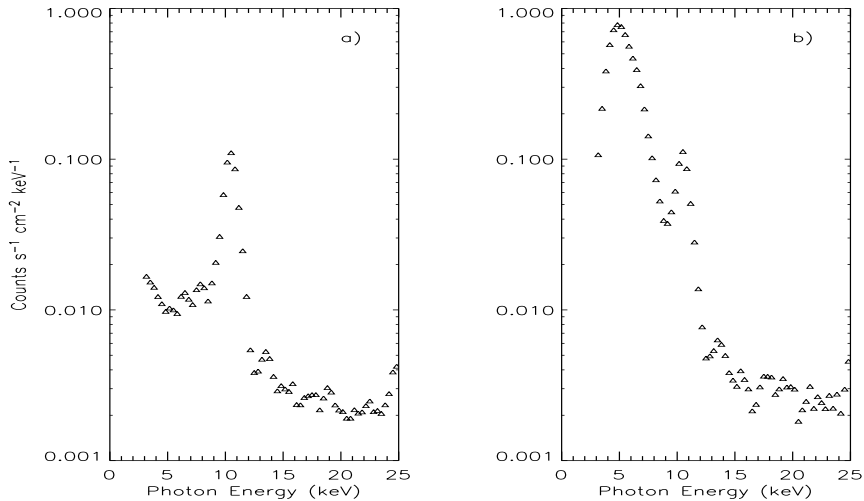}}
\caption{(a) Count spectrum of the night background. (b) Count spectrum 
of the post-flare
background. The spectra are significantly different only for energies 
$E \leq 8$~keV.}\label{plot_back}
\end{figure}
We considered two background spectra: the night
background obtained averaging over few minutes before RHESSI exits the night
for each flare, and the post-flare background determined between 16:55:30 -
16:57:30~UT, well after the decay of any flare.
The night background, which includes all the instrumental contributions, 
represents a sort of lower limit, while
for the post-flare background we point out that the chosen 
interval is more than two hours (and, in RHESSI 
terms, more than an orbit) far from the first flare, and almost half an hour
from the second one. 

In Figure~\ref{plot_back} we compare the two background spectra.
We can see that the post-flare background is at least one
order of magnitude higher than the night background for energy lower 
than 8~keV, while the spectra are
almost identical at higher energies. The
two peaks at around 10.5~keV and 13.5~keV, present in both spectra, 
are probably due to the activation of the germanium detectors.
We then expect the choice of the background will affect the flux spectra
only at low energies.

\subsection{Spectra of the extended flare} \label{spe:pile}
\begin{figure}[t]
\centerline{\includegraphics[width=12cm]{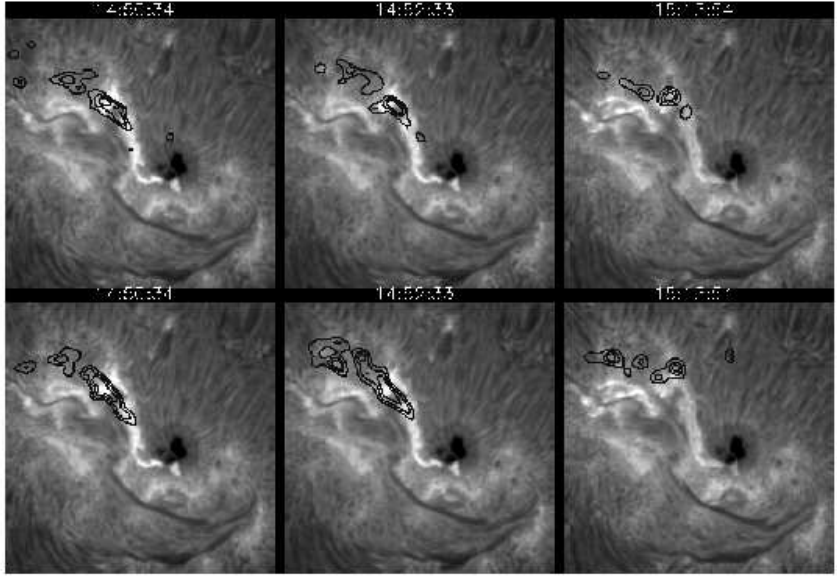}}
\caption{Evolution of the extended flare during the decay phase. 
Images in the H$\alpha$ line
center with overplotted isocontours (40~\%, 60~\%, 80~\% of the peak
intensity) of RHESSI images in the 3~--~6~keV ({\it top}) and 
6~--~12~keV ({\it bottom}), obtained using {\it Pixon} algorithm.}
\label{spe:ima_evol}
\end{figure}

In Figure~\ref{spe:ima_evol} we show the evolution of the extended flare
during its decay phase. The RHESSI images are obtained using {\it Pixon} 
algorithm with detectors 3~--~8  (excluding detector 7 in the 3~--~6~keV
band, see Sect.~\ref{ima:rec:ext} ) and $\Delta t = 60$~s.
 We could select all available collimators because {\it Pixon} decides
which ones to keep and which to discard. 
The precision of the alignment between H$\alpha$ and RHESSI images is
estimated to be around 5\arcsec.
H$\alpha$ images show the decreasing intensity of the ribbon. 
At 14:50:30~UT the RHESSI emission maximum coincides with the 
loop footpoints outlined by the H$\alpha$ west-ribbon and only a
weak emission is present in the northern part of the FOV tracing the
inferred large loop system connecting the two ribbons visible in 
Figure \ref{ima:hasta}.
Enlarging the RHESSI FOV we found  that, from  this time on, no  
X-ray emission is present in the eastern ribbon location. 
During the decay the intensity of H$\alpha$ footpoints decreases as well
as the peak flux of RHESSI images (variations are within a factor 10) and 
the X-ray maximum emission moves  from the 
footpoints to the loop top.

\begin{figure}%[t] 
\centerline{\includegraphics[width=12cm,height=8cm]{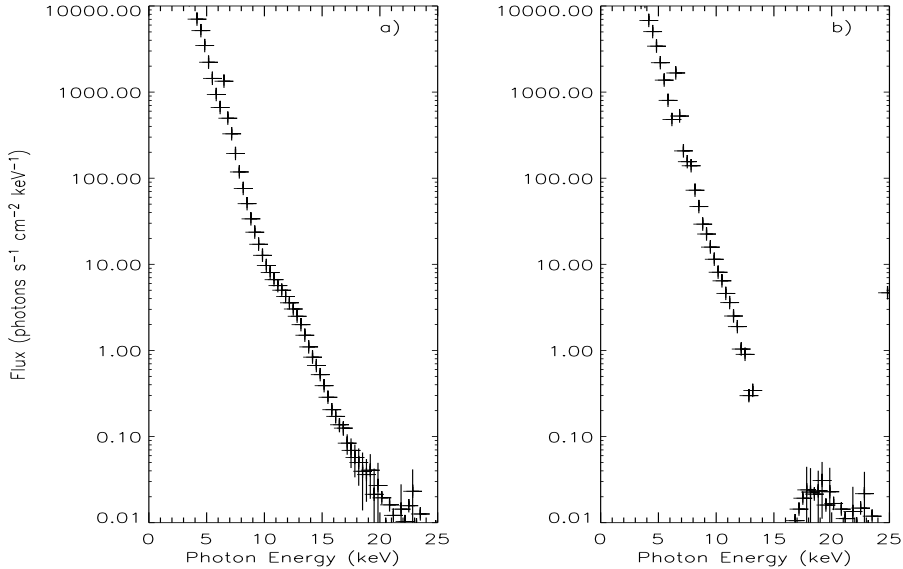}}
\caption{14.50.30~UT  (a) Energy spectrum after
the background subtraction  with no 
pileup correction: the feature between $\approx 10 - 15$~keV is due to
pileup. (b) 
The spectrum after the pileup correction.}\label{pileup}
\end{figure}

At least up to
14:53~UT the counts are $\geq 2000$~count~s$^{-1}$  per detector,  
which is the  estimated threshold for the presence of {\it pileup}
effect when both attenuators are out \cite{SmithBBB}.
When pileup occurs multiple photons are recorded as a single photon
with an energy equal to the sum of energies of the individual photons.
In Figure~\ref{pileup} we show the spectrum before and after
the correction for the pileup. 
While the correction works properly below 12~keV, at higher energies the
spectrum seems overcorrected.
Usually the pileup is present for large flares 
and affects particularly the 20~--~50~keV energy range, and
the error for its correction is estimated of about 20 \% when the spectrum is
flat (Smith, private communication). In our energy range the spectrum is very 
steep and the error could be much higher.
The clear presence of pileup in the spectrum 
prevents us to reconstruct reliable images in the 12 -- 25 keV range.

\begin{table}%[t]
\caption{Parameters obtained fitting the spectra with 3 models during
the decay phase of the extended flare. EM, EM$_{h}$, T and T$_{h}$ are
the emission measure and the temperature, respectively, of the cold and 
hot thermal components, $\gamma$ is the exponent of the power-law for the
non-thermal component. The $\chi^{2}$ value is also given.}
\label{fit_extended}
\begin{tabular}{clcccccc}
\hline
UT&model & EM & T & $\gamma$ & EM$_{h}$ & T$_{h}$ &
$\chi^{2}$ \\

& &  (10$^{48}$ cm$^{-3}$) & MK & & (10$^{48}$ cm$^{-3}$) & MK & \\
\hline
14:50:30 &th. &  1.0 & 12 & & & & 3.7 \\
&th + non th.  & 1.7 & 10 & 10 & & & 2.4 \\
&double th. &  1.7 & 9 & & 0.24 & 14 & 2.4 \\
14:52:30 &th. & 1.2 & 11 & & & & 3.5 \\
&th + non th.  & 2.1 & 10 & 11 & & & 1.9 \\
&double th. &  2.0 & 9 & & 0.10 & 14 & 2.1 \\
15:14:00 &th. & 0.8 & 8 & & & & 1.2 \\ \hline
\end{tabular}
\end{table}

Hence we decided to fit the spectra in the energy range 4~--~10~keV
using the same time intervals of the images.
Contrary to expectations, the background choice does not strongly 
affect the resulting
parameters even in the low energy  part of the spectrum, probably
because of the high counts recorded for this flare.
In table~\ref{fit_extended} we give
the parameters for the 3 models of the photon spectrum, 
obtained subtracting
the post-flare background spectra (see Sec.~\ref{spe:cho_back}).
At 14:50:30~UT and 14:52:30~UT the single thermal model fitted the observed
spectra giving 
a high $\chi^2$ of  about 4 while the other two models fitted the spectra
with low comparable $\chi^2$ values of about 2.
For the non-thermal model we set the cutoff at 7~keV (see 
Sec.~\ref{spe:pha} for the discussion on the best choice of cutoff value) 
and we obtain a very steep power-law index $\gamma \approx 10$,
as found in the decay phase of some RHESSI microflares 
(\opencite{KruckerEtal02}, \opencite{Hannahetal:04}).
For the double thermal model the hot component has a temperature of
about 14~MK, compared with the 9~MK of the `cold' component.

At 15:14~UT instead,  when the footpoint emission is decreased 
and the contribution of the loop top increases, the single thermal model adequately
fits the spectrum (T = 8~MK and EM $ = 0.8 \times 10^{48}~cm^{-3}$) reaching a $\chi^2$ of about 1.
 This is coherent with the scenario of an eruptive two-ribbon flare
characterized by a global magnetic field disruption and processes of energy
release long lasting after the impulsive phase. 
The magnetic reconnection process, responsible for these energy release
episodes, is very fast in the beginning of the flare and slows down in
later phases (\opencite{Carmichael:64}, \opencite{Sturrock:66},
\opencite{Hirayama:74}, \opencite{kopp:76}). We can postulate that 
at 15:14~UT there are no more energy release episodes and thus, when the
footpoints of the loop do not contribute significantly to the total
emission, the plasma emission is only thermal.

\subsection{Spectra of the compact flare} \label{spe:pha}
\begin{figure}%[t]
\centerline{\includegraphics[width=12cm]{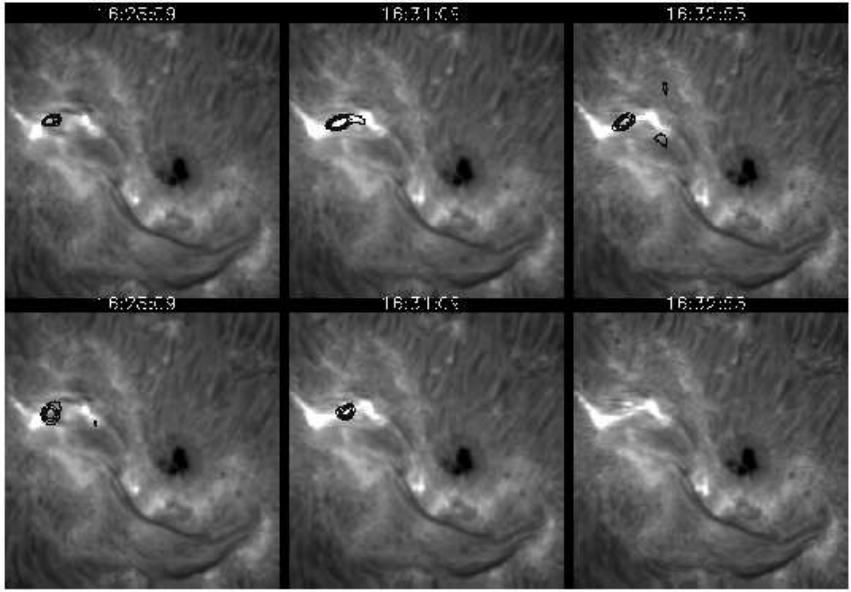}}
\caption{Evolution of the compact flare. 
Images in the H$\alpha$ line
center with overplotted isocontours (40~\%, 60~\%, 80~\% of the peak
intensity that changes within a factor 5) of RHESSI images in 
the 6~--~12~keV ({\it top}) and 
12~--~25~keV ({\it bottom}). The images are obtained using 
{\it Pixon} algorithm with $\Delta$t = 20 s and 60 s, respectively.
At 16:32:53 no reliable image can be reconstructed in the 12~--~25~keV
energy range due to the low counts.}
\label{spe:ima_evol_pha}
\end{figure}
The RHESSI images are obtained using the {\it Pixon} algorithm with
detectors 3~--~7 (the detector 7 is not included in the 3~--~6 keV 
band). In the range 3~--~12~keV we considered $\Delta t = 20$~s,
while in 12~--~25~keV we had to consider $\Delta t = 60$~s (see 
Sec.~\ref{ima:comp}).
In Figure~\ref{spe:ima_evol_pha} we show the evolution of the compact flare
beginning with the rise phase observed by RHESSI (times are indicated in
Figure~\ref{crv:all}). 
The RHESSI contours are very similar in 3~--~6~keV and 6~--~12~keV energy bins
and we show only 6~--~12~keV and 12~--~25~keV contours.
 H$\alpha$ images show that the bright features develop with time along 
the filament
which is well visible before and after the flare (see
Figure~\ref{ima:ubf}). Also the contours of 
RHESSI images in the 6~--~12~keV energy band change with time following the
direction of the filament. In the 12~--~25~keV the peak emission moves 
along the filament and the contours, 
elongated in the North - South direction, probably include the loop
footpoints and 
suggest the presence of low loops crossing the filament.

\begin{figure}%[t]
\centerline{\includegraphics[width=12cm]{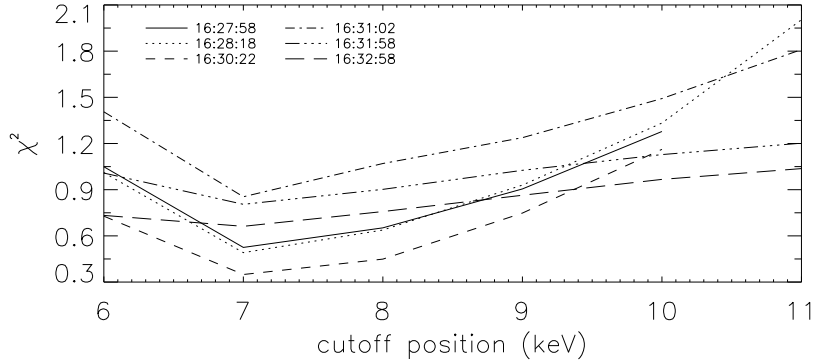}}
\caption{$\chi^2$ vs. the non-thermal energy cutoff value at different times
during the flare. 
}\label{chi}
\end{figure}

We fit the spectra obtained with $\Delta t = 20$~s.
Since no pileup effect is present in this flare, 
the data are reliable in the energy range 4 -- 25 keV. 
We fit the spectra changing the energy range from 4~--~10~keV
to 4~--~25~keV and we see that the found parameters change within 
1 \% while the $\chi^2$ decreases by a factor of two increasing the range
of energy. 

With the single thermal model the fit provides
high $\chi^2$ values ($\approx$ 6~--~8) with high residuals at energies
$\geq$ 10~keV. 
To fit the observed data adding a power-law
non-thermal spectrum at first we analyze the output parameters changing the
position of the energy cutoff from 6~keV to 11~keV.
The found $\gamma$ index can vary up to 15 \% with random residuals.
The resulting $\chi^2$ seems to depend on the energy
cutoff as shown in Figure~\ref{chi} at different times during 
the flare. The minimum value is always reached
with a cutoff between 7~keV and~8 keV, so we adopted 7~keV as the standard
value for the cutoff.
A posteriori, we can justify the use of the data in the 
4~--~10~keV range to fit the spectrum and the setting of the cutoff of
the non-thermal model 
at 7~keV also for the first flare, when the pileup effect makes unreliable
the high energy data. 

\begin{figure}%[ht] 
\centerline{\includegraphics[width=12.5cm]{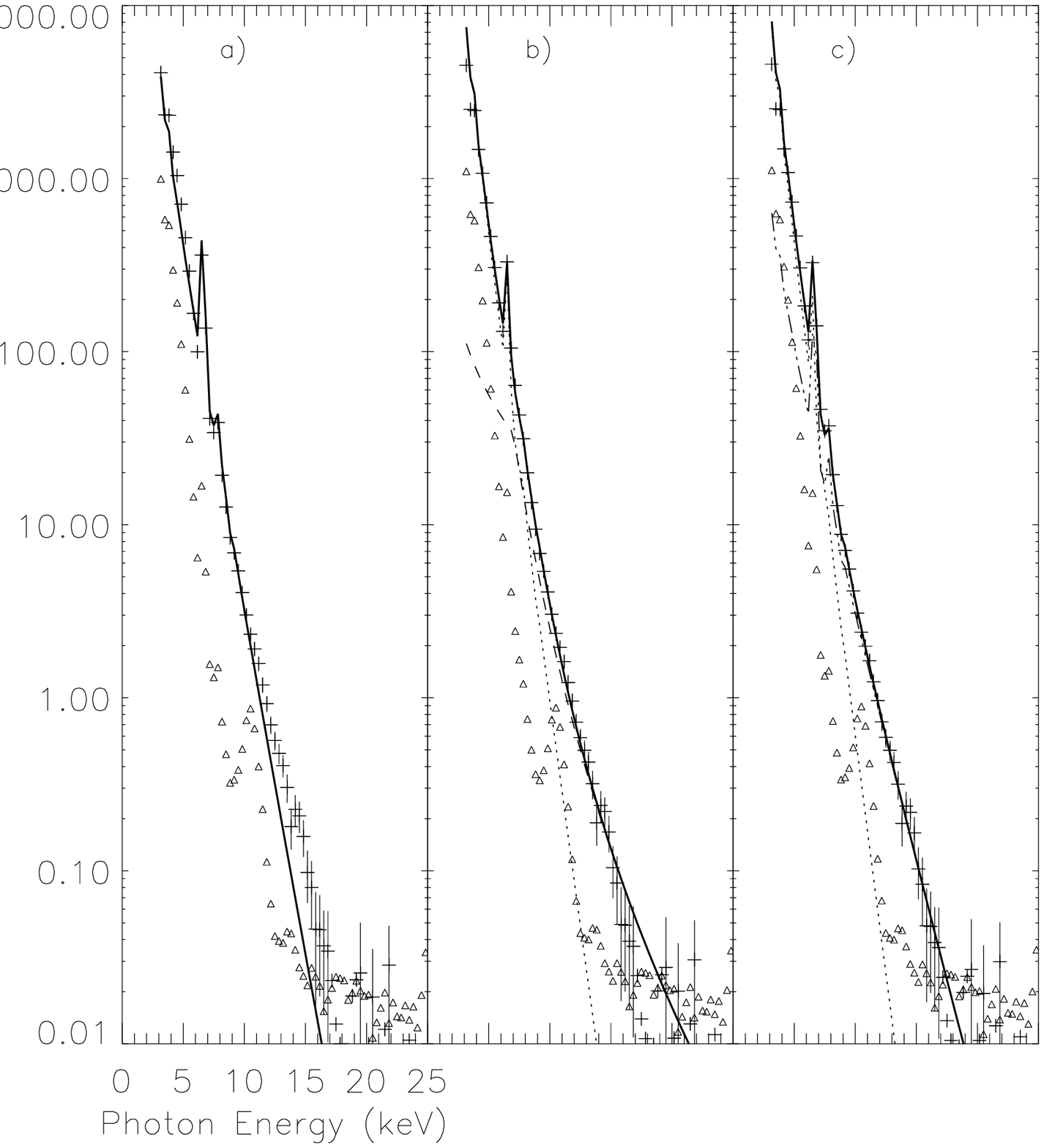}}
\caption{Fitting of the observed spectra (crosses) at 16:31:02~UT with the 3
considered models. The triangles 
indicate the background. The result of the fit is indicated by the 
{\it solid line} and the
thermal component by the {\it dotted line}. (a) 
Single thermal model. (b) Thermal plus non-thermal model ({\it dashed line}).
(c) Double thermal model (the hot
component is indicated by the {\it dot-dashed line}).}\label{spe:models_comp}
\end{figure}

In Figure~\ref{spe:models_comp} we show, as an example, 
the spectra at the time of
the second peak (16:30:52~UT -- 16:31:12~UT) fitted with different models.
The broadened emission feature at around 6.7 keV, corresponding to a group
of emission lines mainly due to  Fe~{\sc XXV} and Fe~{\sc XXVI}, is also taken
into account by the SPEX code in the fitting procedure.
\inlinecite{Phillips04} warned about the fact that SPEX
computations are based on ionization and recombination data that
overestimated the fraction of Fe~{\sc XXV} at T $\approx$ 10~MK.

\begin{table}[t]
\caption{Parameters obtained fitting the spectra with 3 models during
different phases of the compact flare. EM, EM$_{h}$, T and T$_{h}$ are
the emission measure and the temperature, respectively, of the cold and 
hot thermal components, $\gamma$ is the exponent of the power-law for the
non-thermal component. The $\chi^{2}$ value is also given.}
\label{fit_compact}
\begin{tabular}{clcccccc}
\hline
UT&model& EM & T & $\gamma$ & EM$_{h}$ & T$_{h}$ &
$\chi^{2}$ \\
&  & (10$^{48}$ cm$^{-3}$) & (MK) & & (10$^{48}$ cm$^{-3}$) & (MK) &
\\ \hline
16:27:58 &th. & 0.08 & 12 & & & & 5.0 \\
&th. + non th.  & 0.24 & 9.7 & 7.7 & & & 0.6 \\
& double th. &  0.32 & 8.9 & & 6.0 10$^{-3}$ & 20 & 0.5 \\
16:28:18 &th. & 0.11 & 12 & & & & 4.3 \\
&th. + non th.  & 0.26 & 10 & 7.7 & & & 0.5 \\
& double th.  & 0.32 & 9.4 & & 7.5 10$^{-3}$ & 19 & 0.5 \\
16:30:22 &th. & 0.07 & 14 & & & & 7.7 \\
&th. + non th.  & 0.31 & 9.9 & 7.2 & & & 0.5 \\
& double th. &  0.43 & 9.0 & & 7.0 10$^{-3}$ & 21 & 0.4 \\
16:31:02 &th. & 0.09 & 14 & & & & 6.4 \\
&th. + non th.  & 0.33 & 11 & 7.3 & & & 0.9 \\
& double th. &  0.39 & 10 & & 9.1 10$^{-3}$ & 21 & 0.5 \\
16:31:58 &th. & 0.21 & 12 & & & & 2.5 \\
&th. + non th.  & 0.37 & 10 & 8.2 & & & 0.8 \\
& double th. &  0.41 & 9.7 & & 0.010 & 17 & 0.8 \\
16:32:58 &th. & 0.25 & 10 & & & & 1.3 \\
&th. + non th.  & 0.41 & 9.0 & 10.6 & & & 0.6 \\
& double th. &  0.42 & 8.9 & & 4.8 10$^{-3}$ & 16 & 0.7 \\
16:34:02 &th. & 0.27 & 8.7 & & & & 1.2 \\ \hline
\end{tabular}
\end{table}

As expected, the choice of background for this very weak flare 
heavily affects the parameters
of the cold thermal spectrum: the EM obtained subtracting the night background
is approximately twice the EM obtained with the  post-flare background
while the temperature is on the average 0.5~MK lower.
On the contrary the fitting parameters that
characterize the power-law or the hot components 
are not  substantially altered by the background as was also found by
\inlinecite{Qiuetal:04}.
In table~\ref{fit_compact} we give
the parameters obtained for the 3  models taking into account
the post-flare background spectra (see Sec. \ref{spe:cho_back}).
In the rise and peak phases of the flare the non-thermal spectrum 
has a $\gamma$ of about 7
while in the decay phase $\gamma$ increases up to 10, as
in the decay phase of the other flare. 
The high value of $\gamma$ reached when the emission decreases, indicates a
softening of the emission with respect to the peak phase.
If we assume a double thermal model, the hot component is always
present during the flare and has a temperature
T $\approx$ 20~MK (compared to 10~MK of the cold one) and an EM two orders of
magnitude lower than that of the cold component.
The temperature we found is anyway quite lower than the 35~MK indicated for the so
called ``superhot'' component first identified by \inlinecite{Linetal:81} and
afterward 
defined as the component giving rise to the Fe~{\sc XXVI} emission lines
observed in large flares with BCS on Yohkoh
(see e.g. \opencite{Pikeetal:96}).

\subsection{Discussion }
An important result is that, for both flares, only towards the end of the 
decay phase the best fit can be obtained with a thermal photon spectrum while
during the flare the best fit can be obtained, 
with comparable $\chi^2$ values,
adding either a non-thermal component or a hot thermal component.
The two models are equivalent on this statistical base
and it is difficult to select the best model to interpret the observations.
We try to
use the Fe line feature at 6.7~keV, well visible in our observed spectra,
to choose between these two models.
\inlinecite{Phillips04} computed synthetic spectra 
using the CHIANTI code and found that this feature becomes
visible above the continuum only when T $\geq 10$~MK (with an
EM $=10^{49}$~cm$^{-3}$). Its equivalent width gives the best
temperature diagnostic reaching a value of 3~keV for temperature T
$\approx$ 20~MK. 
We measured the equivalent width of this line in the count spectra and
we found a value always $\leq 1.0$~keV that indicates a thermal plasma
with T~$\leq 12$~MK.
We hence believe that the observed spectra could be better interpreted
adding a non-thermal spectrum to a thermal one.

From the fit parameters we can compute the thermal and the non-thermal
contribution to the flare energy budget, 
assuming that the accelerated electrons lose their energy primarily
through Coulomb collisions in the thick target approximation
\cite{Linetal:01}:
\begin{subequation}
\begin{equation}
E_{th} = 3 k_{B} T \sqrt{EM\:V} 
\end{equation}
\begin{equation}
E_{non-th} = 9.5 \times 10^{24}\gamma^2 (\gamma-1) B(\gamma-1/2,3/2) A E_0^{-(\gamma - 1)} \Delta\, t
\end{equation}
\end{subequation}

where $k_B$ is the Boltzmann constant, T and {\it EM} the temperature
and the 
Emission Measure of the thermal part, V the volume of the source,
$B(a,b)$ the standard beta function,
{\it A} the normalization factor of the photon spectrum, 
$E_0$ the energy cutoff, and $\Delta$t the time interval.
T, {\it EM}, $\gamma$ and {\it A} are derived from the fitting. We
estimated
{\it V} from the flare size (see $w_{x,y}$ in Table~\ref{tab:comp}). 
 For the extended flare we have observations only during the decay phase
and we find $E_{th} \approx 1\times 10^{30}$~erg 
and $E_{non-th} \approx 1\times 10^{31}$~erg.
For the compact flare we find that  both $E_{th}$ and $E_{non-th}$ are 
approximately constant
during the flare with $E_{th} \approx 5\times 10^{28}$~erg and 
$E_{non-th} \approx 5\times 10^{29}$~erg. Thus, the non-thermal energy 
is higher than the
thermal energy by about a factor 10  not only in the rise and peak phases
of the second flare but also during the decay phase of both flares. 
The high value of the $\gamma$  index
in this phase, that indicates a softening of the emission, 
balances the low value of the normalization factor of the photon spectrum
{\it A}.
This means that the non-thermal electron energy is the energy supply
for the heating of the thermal plasma.
This result can thus be considered an a posteriori support to the
non-thermal model choice.

\section{Conclusions}
 We use RHESSI data and ground based observations to study two small flares 
(GOES class C1 and C2) developed 
in the same active region
with different characteristics: the first is a two-ribbon extended flare
implying high coronal loops while
the other is a compact two ribbon implying low-lying loops.
This is reflected in RHESSI imaging at different energies that outline
the loop morphology in the two cases.

We analyzed the accuracy of different algorithms
implemented in the RHESSI software to reconstruct the image of the
emitting sources, for energies (between 3 and 12 keV)
that are  particularly important when studying microflares.
We found that all tested algorithms give similar results for the
peak positions and the total flux on the considered FOV, as 
found by \inlinecite{AschwandenEtal04} for higher energies.
The peak flux values and the linear sizes
differ at most by a factor 2. {\it Clean} gives always the lowest peak
and the largest size (does not deconvolve from the PSF of the
instrument) while the other algorithms give very similar
results.  {\it Clean} can give additional problems 
even if the flare structure is simple. In fact for our compact flare it
shows very high negative
residuals out of the main source and then does not conserve the total
flux over the FOV. The available option to
correct for residual map does not help to solve the problem. 
{\it MEM-Sato} always gives the highest value
for H/C, i.e. tends to 
spread more photons out of the source than the other methods, as found
by \inlinecite{AschwandenEtal04}.
 {\it Pixon} and {\it Forward-fit} do converge to similar results. 
Both algorithms have the option to detect and remove the unmodulated flux 
present
in RHESSI data. {\it Forward-fit} finds a unmodulated component that
slightly changes the source parameters for both flares, while in the
case of extended complex flare, {\it Pixon} finds a component that 
strongly  changes the source parameters. 
On the other hand, {\it Forward-fit} is not able to reproduce the real 
complexity of the source, hence
we used {\it Pixon} (without background correction) to reconstruct 
the images of the two flares
during their evolution. In fact, while the reliability of {\it Pixon} is
not surprising,
the most relevant result of our study is that each other method can meet
significant problems, and special care is always required employing
them.
In any case, we outline the importance of comparing different algorithms
when analyzing events in order to establish whether the
reconstructed features are real or not.

We studied the spectral characteristics of our flares by
fitting the count spectra with 3 different models of photon spectrum: 
a single isothermal spectrum, an isothermal plus a non-thermal
spectrum and a third one with two isothermal components. 
The photon spectrum seems to depend on the flare phase. The single
thermal model is adequate to fit the observations only during the very final
decay phase. During the flare, instead,
the single thermal model provides $\chi^2$ values a factor of 10
higher than the one obtained with the other two models.
The non-thermal and the double thermal fits are comparable, and the
choice between them can not rely on statistics.
The measured equivalent width of the Fe line feature at 6.7~keV is low compared
to that expected for a plasma at 20~MK. Although further investigation
is necessary, this may suggest that the hot thermal
component is not relevant to the total emission. 
For the non-thermal spectrum
we determined the energy cutoff value at 7~keV as the one that gives the lowest
$\chi^2$ in the fitting. For the rise and peak phase we found a 
photon spectral index of 7~--~8, 
quite high for standard flares, but not so uncommon in
microflares (\opencite{Kruckeretal:04}, \opencite{Hannahetal:04}), 
while during the decay phase $\gamma$ increases up to 10, indicating a
softening of the emission with respect to the peak phase.
With such a steep spectrum
and a low energy cutoff the non-thermal energy
contribution to the energy budget of the flare 
is always higher than the thermal one and perhaps the non-thermal electrons
provide the energy for heating the thermal plasma through collisional
loss.

As a ``by-product'' of our spectral study,
we found that the pileup effect, usually observed during large flares,
must be taken into account also for 
weak flares and that the correction enabled in the RHESSI software 
can give serious problems for the recovering of the spectra in the 
3~--~25~keV energy
range.  We want to stress that the presence of pileup also affects the
image reconstruction process, yielding unreliable
images, because the photon energies are uncertain.
\begin{acknowledgements}
We thank P. Grigis, D.M. Smith, R.A. Schwartz and G.J. Hurford for their 
precious help about analyzing and interpreting the RHESSI data. 
We would like to thank the anonymous referee for constructive
comments that help us to improve the paper.
We wish to thank the NSO/Sac Peak staff for the time allocation and help
during the observations. NSO is operated by AURA, Inc., under
cooperative agreement with the NSF. We would also like to thank 
R. Falciani and G. Cauzzi who acquired the data at NSO and M. L.
Luoni who provided
us the images obtained at OAFA (El Leoncito, San Juan,
Argentina) in the framework of the German-Argentinean HASTA/MICA
Project, a collaboration of MPE, IAFE, OAFA and MPS.
\end{acknowledgements}

{}
\end{article}
\end{document}